\documentclass[prl,twocolumn]{revtex4}

\usepackage{epsfig}

\begin{document}

\title{Mass Generation and Symmetry Breaking in Chern-Simons Supergravity}

\author{Jorge Alfaro and M\'aximo Ba\~nados}

\affiliation{Departamento de F\'{\i}sica, Pontificia Universidad Cat\'olica de Chile,
Casilla 306, Santiago 22, Chile.
\\ {\tt jalfaro@puc.cl, mbanados@puc.cl }    }

\begin{abstract}

We argue that the quartic fermionic potential of five-dimensional Chern-Simons
supergravity induces spontaneous symmetry breaking, in a phenomenon bearing a close
connection with the Nambu and Jona-Lasinio model.

\end{abstract}

\maketitle

An attractive feature of five-dimensional Chern-Simons supergravity based on the Lie
algebra $SU(2,2|N)$ is that it contains $SU(N)$ gauge fields, fermions and gravity. The
gauge field $\Omega \in SU(2,2|N)$ has the matrix presentation
$$
\Omega = \left( \begin{array}{cc}   W &   \psi^j   \\
                          \bar\psi^i &  {\cal A}^{ij}   \end{array} \right)
$$
where $W\in U(2,2)$, $\psi_i$ are $N$ Dirac spinors, and ${\cal A}^{ij}$ are $U(N)$ gauge
fields.  The action, first considered by Chamseddine \cite{Chamseddine}, is
\begin{eqnarray}
 I[\Omega]  &=& -ik \int \mbox{STr} \left( {1 \over 3}d\Omega d\Omega \Omega + {1 \over 2}
d\Omega \Omega^3  + {1 \over 5} \Omega^5 \right) \nonumber\\
&=& I[W] + I_W[\psi]  \label{I}
\end{eqnarray}
where $I[W]$ is a Chern-Simons theory by itself. Since we shall be interested in the
gravitational and fermionic degrees of freedom we set  Tr$(W)={\cal A}=0$ hereafter.

Note that this theory  has only two (dimensionless) parameters, the number of colors $N$
and the level $k$.  The gravitational equations of motion are
\begin{equation}\label{1}
   {\cal R \wedge R} = - {\delta I_W[\psi]  \over \delta W}
\end{equation}
with ${\cal R} = dW + WW$. On the fermion ground state, it is generally taken from
granted that the right hand side vanishes. The goal of this paper is to study this
assumption in detail.

Chern-Simons gravity was first introduced in \cite{ATW} in three dimensions. It was then
pointed out \cite{Chamseddine} that the same construction can be carried over to five
dimensions. Poincare Chern-Simons supergravities were introduced in \cite{BTrZ}, and
their anti-de Sitter extensions in \cite{TZ}. It was conjectured in \cite{TZ} and
\cite{Horava} that M-theory may be described by eleven-dimensional Chern-Simons
supergravity. This idea has recently been reexamined in \cite{Nastase}. A possible
mechanism relating the field content of standard and Chern-Simons supergravities was
discussed in \cite{B}. The dynamical structure of higher dimensional Chern-Simons
theories and their associated current algebras was studied in \cite{Nair,Dunne,BGH}. See
also \cite{More} for other aspects of higher-dimensional Chern-Simons theories.

Let us start by recalling the meaning of the Chern-Simons level $k$ and its relation to
Planck's scale. In the purely gravitational sector,  the gauge field $W$ is expanded as
\begin{eqnarray}
W = {1 \over 2l} e^a \gamma_a + {1 \over 4} w^{ab} \gamma_{ab}
 \label{W}
\end{eqnarray}
where $e^a$ is identified as the veilbein, and $w^{ab}$ the spin connection. The length
parameter $l$ (AdS radius) is introduced here because $W_\mu$ has dimensions $1/$length
while $e^a_\mu$, related to the spacetime metric by $g_{\mu\nu} = e^a_\mu e^b_\nu
\eta_{ab}$, has dimension zero.

The next step is to find Planck's length, the small scale parameter, in terms of $l$ and
$k$. Expanding $I[W]$ in powers of the curvature tensor we can identify the term $  {k
\over 2l^{3}}\sqrt{-g} R$ and we thus find
\begin{equation}\label{k}
  k   =  {l^3 \over l_p^3}
\end{equation}
where $l_p^3=8\pi G$. The Chern-Simons level $k$ in supergravity then measures the
quotient  ``maximum length"/``minimum length". In the weakly coupled theory with large
$k$ both lengths are far away from each other.

Having introduced the relevant parameters in our theory we shall now set, for notational
simplicity, $l=1$. In this unities, Planck's length and all associated parameters depend
only on $k$.  For later use we only quote Planck's energy
\begin{equation}\label{Ep}
  E_p = {1 \over l_p} = k^{1/3}
\end{equation}

The fermionic term in (\ref{I}) with $N=1$ is  \cite{Chamseddine}
\begin{equation}\label{Ipsi}
  I_W[\psi] =   \int  \left( 2i\, \bar\psi {\cal R} \nabla \psi  + {4i \over k}\bar\psi \psi \bar \psi
\nabla\psi\right)
\end{equation}
where
\begin{eqnarray} \label{nabla}
 \nabla \psi &=& D\psi + {1 \over 2} e^a \gamma_a \: \psi \nonumber\\
{\cal R} &=&  {1 \over 4} \left(R^{ab}+ e^a e^b\right) \gamma_{ab}.
\end{eqnarray}
$D$ is the Lorentz covariant derivative with $D\wedge D = R $ and  $ \nabla \wedge \nabla
=  {\cal R}$. For later convenience, we have reescaled the fermions by $\psi \rightarrow
 (2/k)^{1/2}\psi $.

As in any supergravity theory, the free part of (\ref{Ipsi}) is linearly supersymmetric
around $\psi=0$. The variation of the free part of (\ref{Ipsi}) under
$$
\delta \psi = \nabla \epsilon.
$$
yields $\bar\psi {\cal R} {\cal R} \epsilon$ which is zero when the linear bosonic
equations of motion (\ref{1}) hold.  The key question is whether $\psi=0$ is the true
ground state of the theory or not. If this was not true, the expansion around a non-zero
value $\langle \psi \rangle $ would introduce other quadratic terms coming form the
potential. Then, the linear transformations would not cancel on the background (\ref{1})
and the ground state would not be supersymmetric.

We shall argue here that this in fact occurs in the generic situation. In particular we
prove that the fermion condensate
\begin{equation}\label{cond}
  \langle  \bar\psi \gamma^{ab} \psi \rangle \neq 0
\end{equation}
is different from zero. Our analysis has a close analogy with the Nambu and Jona-Lasinio
model \cite{NJL}, although we follow the auxiliary field formalism of Gross and Neveu
\cite{GN}.

The occurrence of (\ref{cond}) follows from the following observation. Recalling the
expressions (\ref{nabla}) for $\nabla$, and applying the five-dimensional Fierzing
identity \cite{Cremmer} it is a simple exercise to express the interaction term as
$$
\label{pot1} {4i \over k}\bar\psi \psi \bar \psi \nabla\psi = {1 \over k}\bar\psi \psi
\bar \psi D\psi - {1 \over 4k} \epsilon_{abcde} e^a \bar\psi \gamma^{bc} \psi\:\bar\psi
\gamma^{de} \psi.
$$
Both terms are perturbations of the same order in powers of $k$. The second term has the
structure $(\bar\psi \gamma^{ab} \psi)^2$, similar to that arising in the Gross-Neveu
model, and we shall then focus on it. It is important to stress, however, that the first
term could be relevant and it may turn on other operators like $\langle \bar\psi
\gamma^{ab} D\psi \rangle$;  we shall study this possibility elsewhere.

Keeping only the second term we rewrite the interaction introducing an auxiliary field
$\sigma^{ab}$,
$$
 -{1 \over 4k}\bar\psi \psi \bar \psi \nabla\psi \rightarrow {k \over 4 }
\epsilon_{abcde} e^a \sigma^{bc}  \sigma^{de} -{1 \over 2 }  \epsilon_{abcde} e^a
\sigma^{bc}\: \bar\psi \gamma^{de} \psi  \label{pot2}
$$
 whose equation of motion is
$$ \sigma^{ab} = {1 \over k} \bar \psi \gamma^{ab} \psi.
$$
The weak coupling, $k\rightarrow\infty$, fermionic action we consider is then,
$$
I_W[\psi,\sigma] = \int\left[ \bar\psi \left( 2i\, {\cal R} \nabla  - { \slash\!\!\!
\sigma \over 2} \right) \psi + {k \over 4 }  \epsilon_{abcde} e^a \sigma^{bc}
\sigma^{de}\right]
$$
where the 3-form $\slash\!\!\! \sigma$ is defined as $\slash\!\!\! \sigma :=
\epsilon_{abcde} e^a \sigma^{bc} \gamma^{de}$. It is now evident that if $\sigma\neq 0$,
this action is not linearly supersymmetric under $\delta \psi = \nabla\epsilon$.

The effective potential $U(\sigma)$ governing the values of $\sigma$ is defined as
$$
 e^{i\int (-U(\sigma))} = \int D\psi D\bar\psi \: e^{iI_{W}[\psi,\sigma]}
$$
and we obtain
$$ U(\sigma^{ab}) = -{k \over 4} \epsilon_{abcde} e^a \sigma^{bc} \sigma^{de} +i
\log\det\left( 2i\, {\cal R} \nabla - {\slash\!\!\! \sigma \over 2}\right).
$$

In a semiclassical approximation, the value of $\sigma$ is given by the minimum of the
effective potential $U$. If this minimum is not zero, then supersymmetry is broken. We
already see that in the limit of large $k$ the solution is in fact $\sigma^{ab}=0$.  We
now study the contribution from the first quantum correction.

Although it is nice to have general background independent formula for $U$, the actual
computation of the determinant is complicated because the operator ${\cal R}\nabla $ is
non-minimal and the standard heat kernel formulae cannot be applied in a straightforward
way.

In order to get an idea into the structure of the determinant we consider the large $l$
regime. In this regime, spacetime is approximately flat and we shall compute the explicit
value of $U$ in that case. Specifically we consider fields $W$ which are slowly varying,
while $\psi$ is fast varying, as compared to $l$. In this regime we approximate
\begin{eqnarray}
R^{ab} + e^a e^b &\approx& e^a e^b \nonumber\\
 D\psi + {1 \over 2} e \psi &\approx& d\psi
\end{eqnarray}
Rather than computing the effective potential directly from the above formula, it is
convenient to rewrite the fermionic action in flat space.  We first note that if the
2-form $\sigma^{ab}$ can have non-zero values, Poincare invariance dictates its general
form,
\begin{equation}\label{sigma}
  \sigma^{ab} = {m \over 6}\: e^a \wedge e^b,
\end{equation}
where $m$ is a constant which can be interpreted as the fermion mass.  The action
$I_W[\psi,\sigma]$ reduced to flat space becomes,
$$
I[\psi]
 = \int \left[\bar\psi_\mu \left(  \gamma^{\mu\rho\sigma} \partial_\rho -
 m \: \gamma^{\mu\sigma} \right)\psi_\sigma - {5 \over 6}k \: m^2 \right]
$$
Of course, if $m=0$ $I[\psi]$ is linearly supersymmetric under $\delta\psi_\mu =
\partial_\mu\epsilon$, and confirms that this action is the flat space analogous to
$I_W[\psi,\sigma]$.

The question we would like to ask is whether or not the true minimum of the effective
potential is $m = 0$.

The determinant can now be computed by direct calculation.  We first compute the
determinant of the Dirac matrices and obtain
\begin{equation}\label{det1}
  \det (\gamma^{\mu\sigma\rho} \partial_\rho - m\, \gamma^{\mu\sigma}) = 2^8 m^8 (m^2+
\partial^\mu\partial_\mu)^6
\end{equation}
The pole at $m=0$ is expected because the action is gauge invariant at this point, and
the integration over fermions would require gauge fixing. This pole is however cancelled
by the determinant of the second class constraints present in the fermionic action. Let
us pause to explain this point.

In a theory with second class constraints $G_A\approx 0$ with $\det \{G_A,G_B\}\neq 0$,
the functional measure contains the factor $\det^{\pm {1 \over 2}} \{G_A,G_B\}$ where the
plus/minus sign corresponds to boson/fermions (see \cite{HT}).

The flat space fermionic action in a 4+1 decomposition reads:
\begin{eqnarray}
{\cal L} &=& -\psi_i^\dagger \gamma^{ij} \dot{\psi}_j + \bar\psi_i (
\gamma^{ijk}\partial_j - m\gamma^{ik})  \psi_k - {5 \over 6}k \: m^2
 \nonumber\\
& &- \psi_0^\dagger (\gamma^{ij} \partial_j - m\: \gamma^i) \psi_i + \psi_i^\dagger
(-\gamma^{ij}\overleftarrow{\partial_j} - m\: \gamma^{i} )\psi_0 \nonumber
\end{eqnarray}
>From here we identify the Poisson bracket
\begin{equation}
  \{ \psi_{ i}^{\ \alpha}(x), \psi^\dagger_{j\ \beta}(y)\} = \left(\hat{\gamma}_{ij}\right)^\alpha_{\
\beta} \delta^{(4)}(x,y) \nonumber
\end{equation}
where $\hat{\gamma}_{ij} \gamma^{jk} =\delta^k_i$, and the two constraints,
\begin{equation}
   G := (\gamma^{ij} \partial_j - m \: \gamma^i) \psi_i, \ \ \ \
G^\dagger :=\psi_i^\dagger (-\gamma^{ij}\overleftarrow{\partial_j} - m\: \gamma^{i} )
\end{equation}
whose Poisson bracket
$$
\{ G(x),G^\dagger(y) \} = {4 \over 3}m^2\: \delta^{(4)}(x,y),
$$
is invertible, as claimed. These constraints become first class in the limit $m
\rightarrow 0$ and generate the supersymmetry transformations. For $m\neq 0$ they are
second class.

Since both $G$ and $G^\dagger$ carry spinor indices, there are in total 8 constraints
$G_A$ ($A=1..8$), and since their Poisson bracket scales as $m^2$ we have
$$
\det \{G_A,G_B\}^{\! -{1 \over 2}}  \sim m^{-8}.
$$
This factor in the fermion measure cancels the pole in (\ref{det1}) exactly. This
cancellation does not mean that we can extrapolate our results all the way to $m = 0$. It
only means that our calculation is correct no matter how small $m$ is; at the exact value
$m \equiv 0$ the whole theory is different, gauge fixing is necessary, and (finite)
discontinuities are actually expected to arise \cite{discont}.

The final expression for the effective potential in momentum representation and Euclidean
space is then
$$
U(m) =  {5k \over 6} m^2 - 6\int_\Lambda {d^5p \over (2\pi)^5} \log\left(1+ {m^2\over
p^2}\right).
$$
where we have chosen $U(0)=0$ and $\Lambda$ is an $SO(5)$-invariant UV cutoff. Before
going any further, let us recall that the fermionic quartic interaction we have
considered here is not renormalizable in five dimensions. As a consequence, expanding the
potential $U(m)$ into positive powers of $\Lambda$ one finds a term $m^4 \Lambda$ that
cannot be absorbed by a redefinition of the coupling $k$. We shall proceed by identifying
the physical parameters of the theory and write the cutoff in terms of them. Not
unexpectedly, we find that $\Lambda$ must be of the order of Planck's energy.

The potential $U(m)$ is clearly stable since, for a given cutoff,
$U(m\rightarrow\pm\infty) \rightarrow +\infty $, and there are no other poles. The next
question is whether symmetry breaking takes place. Taking the derivative of $U$ with
respect to $m$ one finds the equation,
\begin{equation}\label{gap1}
  {5k m \over 3} = \int_\Lambda{d^5p \over (2\pi)^5}  { 12 m\over m^2+p^2},
\end{equation}
describing the minima of $U$. One solution to this equation is $m=0$ corresponding to the
unbroken phase. We would like to know if there are other solutions with $m\neq 0$, and
less energy. In the following discussion it will be convenient to redefine the
Chern-Simons coupling $k$ as,
\begin{equation}\label{k'}
 k = {k'^3 \over 5\pi^3}.
\end{equation}
As it happens for all dimensions greater than two \cite{NJL,RWP}, Eq. (\ref{gap1}) has a
two-phase structure depending on the values of $k'/\Lambda$. In fact evaluating the
integral in Eq. (\ref{gap1}) we see that its non-zero solutions  must satisfy
\begin{eqnarray}
  {k'^3 \over \Lambda^3} &=&   1 - 3\:{m^2 \over \Lambda^2} + 3 \:{m^3 \over \Lambda^3}
\arctan{\Lambda \over m}.
 \label{gap2}
\end{eqnarray}
This equation is plotted in Fig.1.  The left hand side is represented by straight lines,
and the right hand side by the curve starting at 1, for $m=0$, and going monotonically to
zero as $m \rightarrow \infty$.  The intersections define non-zero solutions. The graph
is symmetrical under $m \rightarrow -m$.

\begin{figure}
\centerline{ \epsfig{file=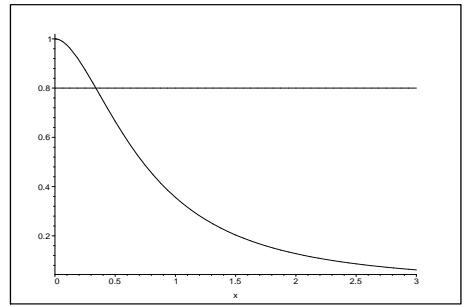,width=4cm,height=6cm,angle=-90}} \caption{Graphical
analysis of Eq. (\ref{gap2}). The horizontal lines represent different values for
$k'^3/\Lambda^3$. Non-zero solutions (intersections) exist only for $k'^3/\Lambda^3 <1$.
The value of $m$ at which the intersection occurs increases as $k'^3/\Lambda^3
\rightarrow 0$. }
\end{figure}

We first note that the intersection occurs only if
\begin{equation}\label{gap3}
  {k' \over \Lambda} <1.
\end{equation}
If this condition is not fulfilled, the only solution to (\ref{gap1}) is $m=0$, and the
symmetry is not broken.

From (\ref{k'}) and (\ref{Ep}) we see that $k'=5^{1/3}\pi\, E_p$ where $E_p$ is Planck's
energy. Since the cutoff is the largest energy scale we conclude that the physical regime
is in fact the broken phase. Eq. (\ref{gap2}) is satisfied, and non-zero solutions for
$m$ exist. We shall see, however, that $\Lambda$ should not be too big.

Assume then that a non-zero solution to (\ref{gap2}) exists, and let us call it $M$. We
identify $M$ with the physical mass of the fermions. On physical grounds we expect
$M<<\Lambda$, and hence we focus on the region near the origin of Fig.1.  In this region
we can approximate (\ref{gap2}) and $M$ satisfies,
\begin{eqnarray}
  k'^3  &=&  \Lambda^3 - 3 \Lambda M^2.
\end{eqnarray}

Now, since $M$ is the physical mass ``fixed by experiments" it is natural to use this
equation to eliminate the cutoff in terms of the two physical quantities $M$ and $k'$. To
first order in $M^2/k'^2$ we find
\begin{equation}\label{gap4}
  \Lambda  = k' \left(1  + {M^2 \over k'^2}+\cdots \right).
\end{equation}
Now we replace this value of the cutoff in the original potential and find the leading
contribution in the large $k'$ expansion
\begin{equation}\label{U2}
  U(m) = {k' \over 4\pi^3} ( m^4 - 2M^2 m^2) + {\cal O}(1).
\end{equation}
This potential does not depend on the cutoff and exhibit symmetry breaking for all
positive values of $M^2$. By construction the value of $m$ at the minimum is $M$, and
$$
U(M) = - {k' \over 4\pi^3} M^4 < U(0).
$$
Hence $m=M$ and not $m=0$ represents the true ground state.

It is interesting now to observe that $\Lambda \sim k' \sim k^{1/3}$ and, in view of
(\ref{Ep}), we find the expected result $\Lambda \sim E_p$ (plus small corrections given
by the fermions mass). Eqns. (\ref{gap2}) and (\ref{Ep}) tell us that the fermion mass
will satisfy $m<<\Lambda$, provided $\Lambda$ is of the order of $E_p$. If the cutoff is
taken all the way to infinity then $m$ diverges as well.

We would like to end with some comments and future prospects.  We have shown that the
effective potential governing the vacuum expectation value of the operator $\sigma^{ab}
=(1/k) \langle \bar\psi \gamma^{ab} \psi\rangle$ does exhibit symmetry breaking. We have
only computed the value for the potential on a flat background: it would be very
interesting to find $U$ for an arbitrary background. This would allow the study of the
back reaction from the fermions fields to the geometry, and would  be particularly
relevant in view of the results of \cite{B}. In fact the original motivation for this
calculation was to study the possibility of non-trivial vacuum sources in (\ref{1})
coming from the fermions, but in order to analyze this point properly we need the
effective potential on a general field $W$.  Finally, it would also be interesting to
study the effect of the interaction term $(1/k)\bar\psi \psi \bar\psi D\psi$ that we have
discarded in the large $k$ limit.

The work of M.B. was partially funded by GRANTS \# 1020832 from FONDECYT (CHILE). MB
would like to thank Jorge Zanelli for his warm hospitality during a visit to CECS,
Valdivia, where part of this work was done. The work of J.A. is partially supported by
FONDECYT \# 1010967. He thanks the hospitality of LPTENS (Paris), Universidad de
Barcelona and Universidad Aut\'onoma de Madrid. M.B. and J.A. acknowledge financial
support from the Ecos(France)-Conicyt(Chile) project\# C01E05.


\begin{thebibliography}{10}

\bibitem{Chamseddine}
A.~H.~Chamseddine, ``Topological Gravity And Supergravity In Various Dimensions,'' Nucl.\
Phys.\ B {\bf 346}, 213 (1990).


\bibitem{ATW}
A.~Achucarro and P.~K.~Townsend, ``A Chern-Simons Action For Three-Dimensional Anti-De
Sitter Supergravity Theories,'' Phys.\ Lett.\ B {\bf 180}, 89 (1986).
E.~Witten, ``(2+1)-Dimensional Gravity As An Exactly Soluble System,'' Nucl.\ Phys.\ B
{\bf 311}, 46 (1988).

\bibitem{BTrZ}
M.~Ba\~nados, R.~Troncoso and J.~Zanelli, ``Higher dimensional Chern-Simons
supergravity,'' Phys.\ Rev.\ D {\bf 54}, 2605 (1996) [arXiv:gr-qc/9601003].

\bibitem{TZ}
R.~Troncoso and J.~Zanelli, ``New gauge supergravity in seven and eleven dimensions,''
Phys.\ Rev.\ D {\bf 58}, 101703 (1998) [arXiv:hep-th/9710180].

\bibitem{Horava}
P.~Horava, ``M-theory as a holographic field theory,'' Phys.\ Rev.\ D {\bf 59}, 046004
(1999) [arXiv:hep-th/9712130].


\bibitem{Nastase}
H.~Nastase,
arXiv:hep-th/0306269.

\bibitem{B}
M.~Ba\~nados, ``Charged solutions in 5d Chern-Simons supergravity,'' Phys.\ Rev.\ D {\bf
65}, 044014 (2002) [arXiv:hep-th/0109031].
M.~Ba\~nados, ``The linear spectrum of OSp(32$|$1) Chern-Simons supergravity in eleven
dimensions,'' Phys.\ Rev.\ Lett.\  {\bf 88}, 031301 (2002) [arXiv:hep-th/0107214].

\bibitem{Nair}
V.~P.~Nair and J.~Schiff,
Nucl.\ Phys.\ B {\bf 371}, 329 (1992).

\bibitem{Dunne}
G.~V.~Dunne and C.~A.~Trugenberger,
Annals Phys.\  {\bf 204}, 281 (1990).

\bibitem{BGH}
M.~Banados, L.~J.~Garay and M.~Henneaux,
Nucl.\ Phys.\ B {\bf 476}, 611 (1996) [arXiv:hep-th/9605159].


\bibitem{More}
M.~Hassaine, R.~Troncoso and J.~Zanelli, ``Eleven-dimensional supergravity as a gauge
theory for the M-algebra,'' arXiv:hep-th/0306258.
P.~Mora,
Nucl.\ Phys.\ B {\bf 594}, 229 (2001) [arXiv:hep-th/0008180].
J.~Gegenberg and G.~Kunstatter,
Phys.\ Lett.\ B {\bf 478}, 327 (2000) [arXiv:hep-th/9905228].


\bibitem{NJL}
Y.~Nambu and G.~Jona-Lasinio, ``Dynamical Model Of Elementary Particles Based On An
Analogy With  Superconductivity. I,'' Phys.\ Rev.\  {\bf 122}, 345 (1961).

\bibitem{GN}
D.~J.~Gross and A.~Neveu, ``Dynamical Symmetry Breaking In Asymptotically Free Field
Theories,'' Phys.\ Rev.\ D {\bf 10}, 3235 (1974).

\bibitem{Cremmer}
E.~Cremmer, ``Supergravities In 5 Dimensions,'' LPTENS 80/17
{\it Invited paper at the Nuffield Gravity Workshop, Cambridge, Eng., Jun 22 - Jul 12,
1980}

\bibitem{HT}
M.~Henneaux and C.~Teitelboim, ``Quantization Of Gauge Systems,''
 Princeton, USA: Univ. Pr. (1992).

\bibitem{discont}
S.~Deser, J.~H.~Kay and K.~S.~Stelle, ``Hamiltonian Formulation Of Supergravity,'' Phys.\
Rev.\ D {\bf 16}, 2448 (1977).
S.~Deser and A.~Waldron, ``(Dis)continuities of massless limits in spin 3/2-mediated
interactions  and cosmological supergravity,'' Phys.\ Lett.\ B {\bf 501}, 134 (2001)
[arXiv:hep-th/0012014].
M.~J.~Duff, J.~T.~Liu and H.~Sati, ``Quantum discontinuity for massive spin 3/2 with a
Lambda term,'' arXiv:hep-th/0211183.

\bibitem{RWP}
B.~Rosenstein, B.~Warr and S.~H.~Park,
Phys.\ Rept.\  {\bf 205}, 59 (1991).



\end{thebibliography}
\end{document}